\documentclass[final,3p,times,twocolumn]{elsarticle}

\usepackage{graphics}
\usepackage{graphicx}
\usepackage{epsfig}

\usepackage{amssymb}
\usepackage{amsthm}

\biboptions{square,sort&compress}

\begin{document}

\begin{frontmatter}



\title{Study of the threshold line between macroscopic and bulk behaviors for homogeneous type II superconductors}


\author[1] {A.\ Presotto}
\author[2,3] {E.\ Sardella}
\author[1] {R.\ Zadorosny}

\address[1] {Grupo de Desenvolvimento e Aplica\c{c}\~{o}es de Materiais, Faculdade de Engenharia de Ilha Solteira, Univ Estadual Paulista - UNESP, Departamento de F\'{i}sica e Qu\'{i}mica, Caixa Postal 31, CEP 15385-000, Ilha Solteira, SP, Brazil}

\address[2] {Faculdade de Ci\^{e}ncias, Univ Estadual Paulista - UNESP, Departamento de F\'{i}sica, Caixa Postal 473, CEP 17033-360, Bauru, SP, Brazil}

\address[3] {UNESP-Universidade Estadual Paulista, IPMet-Instituto de Pesquisas Metereol\'ogicas, CEP 17048-699 Bauru, SP, Brazil}

\begin{abstract}

In this work we solved the time dependent Ginzburg-Landau equations to simulate homogeneous superconducting samples with square geometry for several lateral sizes. As a result of such simulations we notice that in the Meissner state, when the vortices do not penetrate the superconductor, the response of small samples are not coincident with that expected for the bulk ones, i.e., $4\pi M=-H$. Thus, we focused our analyzes on the way which the $M(H)$ curves approximate from the characteristic curve of bulk superconductors. With such study, we built a diagram of the size of the sample as a function of the temperature which indicates a threshold line between macroscopic and bulk behaviors.

\end{abstract}


\end{frontmatter}


\section{Introduction}
\label{Intro}

The advances in nanofabrication techniques which occurred on the last decades stimulated the production and, consequently, experimental and theoretical studies of superconducting samples with sizes of the order of their fundamental lengths, i.e., $\lambda(T)$ and $\xi(T)$. In such materials the superconducting properties and the vortex dynamics are hugely affected by confinement effects. As a consequence, multi and giant vortex states takes place \cite{schweigert,melnikov, mertelj,baelus,geim1,chibotaru,sardella1,sardella4,grigorieva,moshchalkov,sardella2,pascolati,yampolskii,zhao,benxu,cren,kanda,golubovic,kanda2,milosevic} as well as the coexistence of vortex and antivortex pairs.\cite{chibotaru} Such systems also present others exotic behaviors as non-quantized vortex penetration \cite{geim3} and the arrangement of the vortex lattice in several geometries which follow the symmetry of the samples \cite{schweigert,baelus,sardella4,buzdin,palacios,baelus2,cabral,zad2,misko,sardella3,misko2,zhao2,zad1}.

Recently, theoretical studies with mesoscopic superconductors of type I shown that an applied current in a slab induced the penetration and annihilation of single quantized vortex in the intermediate state \cite{berdiyorov}.  The confinement effects can also induced the suppression of the intermediate state and drastically changed the size and temperature dependence of the critical fields of such materials \cite{muller}.

It is interesting to note that, in all cited works, the mesoscopic superconductors are treated, generically, as  materials of reduced dimensions of the order of $\lambda(T)$ or $\xi(T)$. However, no much attention is done on the real sizes for which a sample could be defined as a mesoscopic specimen. Thus, it is worth to emphasize that the knowledge of the relation between size and superconducting behavior is very important to guide the researchers in their theoretical and experimental studies. Recently, Connolly {\it et al.} \cite{connolly} published a work where they used a criterion based on the competition between the Abrikosov vortex lattice and a shell-like ordering to define a meso-to-macroscopic crossover of a superconducting disk. With the same purpose, the authors of the Ref.~\cite{zad2} proposed the existence of a threshold line between mesoscopic and macroscopic superconducting behaviors. For the mesoscopic-like behavior, the confinement effects are strong enough to induce a crossover of the the vortex lattice and the vortices arrangement follows the symmetry of the sample. Nevertheless, the macroscopic behavior is mainly characterized by some volumetric properties like the value of the upper critical field $H_{c2}(T)$ and the presence of the hexagonal vortex lattice. However, in this state the outer vortices are influenced by the surface and are arranged in a kind of shell.

A possible question that should arise from such analysis is about the typical sizes for which the surface effects could be neglected. Thus, in the present work we determined a possible threshold line between the macroscopic and bulk behaviors. This last one have been defined as the samples for which the influence of the surface on the vortex dynamics could be neglect. In this way, the outline of our work is as follow. First, in section \ref{Formalism},
we provide an overview of the theoretical formalism used to run the simulations. Next, in section \ref{Corss} we describe some definitions used in this paper and an overview of a previous work which was the motivation for the present one. In the remainder sections \ref{Results} and \ref{Conc} we discuss our results and the criteria used to obtain the crossover line between macro-to-bulk behaviors and present our conclusions.

\section{Theoretical Formalism}
\label{Formalism}
The phenomenological theory developed by Ginzburg and Landau (GL for short) \cite{GL} is a very important tool to study the behavior of type I and type II superconductors. In such theory, the superconducting state is described by a complex order parameter $\psi$, for which the physical quantity $|\psi|^2$ represents the density of superconducting carriers, i.e., the Cooper pairs, and the vector potential ${\bf A}$ which is related with the local magnetic field by ${\bf h}=\mbox{\boldmath
$\nabla$}\times{\bf A}$. The Ginzburg-Landau equations in their time-dependent form are expressed by \cite{schimid}

\begin{eqnarray}
\left ( \frac{\partial}{\partial t} + i\Phi\right ) \psi& = & -\left
(-i\mbox{\boldmath
$\nabla$}-{\bf A} \right )^2\psi \nonumber \\
& & +(1-T)\psi(1-|\psi|^2)\;,\nonumber \\
\beta\left ( \frac{\partial{\bf A}}{\partial t}+\mbox{\boldmath
$\nabla$}\Phi \right ) & = & {\bf J}_s-\kappa^2\mbox{\boldmath
$\nabla$}\times{\bf h}\;,\label{tdgleq}
\end{eqnarray}
where ${\bf J}_s=(1-T)\Re\left [ \psi^{*}\left ( -i\mbox{\boldmath
$\nabla$}-{\bf A} \right )\psi \right ]$ is the supercurrent density,
and $\Phi$ is the scalar potential; these two equations are commonly referred to as time dependent Ginzburg-Landau equations (TDGL for short). Thus, the time evolution of a superconducting system and, consequently, the evolution of the vortices even in non-stationary states, could be followed. However, for our purposes, we will use the TDGL equations just as a relaxation method to achieve the stationary state. This is only a matter of convenience, since we could solve the GL equations by other means. For example, we could solve then by finite elements methods (see for instance Ref. ~\cite{qdu}.

Here, the distances are measured in units of the coherence length
at zero temperature $\xi(0)$; the magnetic field is in units
of the zero temperature upper critical field $H_{c2}(0)$;
the temperature $T$ is in units of the critical
temperature $T_c$; the time is in units of the characteristic
time $t_0=\pi\hbar/8k_BT_c$; $\kappa$ is the Ginzburg-Landau parameter;
$\beta$ is the relaxation time of $\bf A$, related
to the conductivity. Rigorously speaking, the Ginzburg-Landau theory is applicable only for temperatures close to $T_c$. However, as we are interested in a general feature of a threshold line between macro-to-bulk superconducting behaviors, we have adopted a linear
dependence with respect to the temperature
for the phenomenological parameters in
the Ginzburg-Landau theory, i.e., $H_{c2}(T)=H_{c2}(0)(1-T)$.\footnote{We have chosen the simplest model for the temperature dependence of the physical quantities. Other better choices which are valid for $T$ well below the critical temperature do not invalidate the present investigation, since our main aim is to show that there is a length scale for which we have a meso-to-bulk crossover, no matter what is temperature dependence of $H_{c2}(T)$.} It is interesting to emphasize that the Ginzburg-Landau theory was proven to give good qualitative results in mesoscopic superconductors even at low temperature, despite the microscopic derivation of the Ginzburg-Landau equations being valid only for T very close to $T_c$ \cite{geurts,geurts2}. For better quantitative comparisons at low temperatures, one should employ Boguliobov-deGenes\cite{bogoliubov,degennes,caroli,kalllin,kato,huo}, Eilenberger\cite{eilenberger,klien,ullah,belova,ichioka}, or recently developed Extended GL model\cite{vagov}.


In this work we solved the TDGL equations for very long cylinders of square cross section and with several lateral sizes, expressed by $L/\xi(0)$, as described in references \cite{zad1,zad2}. Those equations were discretized following the link variables method as developed by Gropp and coworkers \cite{gropp}. It is interesting to emphasize that the TDGL equations, even in their discretized form, are gauge invariant under the transformations
$\psi^{\prime}=\psi e^{i\chi}$,
${\bf A}^{\prime}={\bf A}+\mbox{\boldmath
$\nabla$}\chi$, $\Phi^{\prime}=\Phi-\partial\chi/\partial t$,
where $\psi$ is the order parameter, ${\bf A}$ is the vector potential, $\Phi$ is the scalar potential and $\chi$ is a scalar function. In this study we chose the zero-scalar potential gauge, that is, $\Phi^\prime=0$, at all times and positions.

The simulations were carried out for a type II superconductor with $\kappa=5$, and we focused the analyzes on the Meissner state, i.e., the region of small intensities of magnetic fields which were applied along the cylinder axis. The field was incremented in steps of $\Delta H=10^{-3}$. Although the TDGL equations can provide all the metastable states of a fixed field, as stressed previously, in the present work we studied only the stationary states. We used the value of $\beta = 1$. This choice has no influence on the final configuration of the stationary state since it affects only the time steps to achieve the steady state.\cite{buscaglia}

\section{Crossover Criteria}
\label{Corss}

In order to facilitate the discussion of our results, in this work we have used the following terminology. First, by \textit{mesoscopic} we mean a superconductor of dimensions such that the vortex lattice is mostly influenced by the geometry of the sample. In addition, in the mixed state, the magnetization is not a smooth function of the applied field; it has a series of jumps which indicate the nucleation of one or more vortices. Second, according to references \cite{zad1,zad2}, as the size of the sample is increased, there is a length scale above which deep inside the superconductor, the vortex lattice is not perturbed by the surface effects, although they are still present. In this regime, which we denote by \textit{macroscopic}, the vortices are arranged nearly as a triangular lattice, except near the surface where there are some distortions. Also, in the mixed state, the height of the jumps in the magnetization curves are very small so that it approaches to a continuous line. Finally, by \textit{bulk} superconductors we mean those with an infinite size such that the vortex configuration is a perfect triangular lattice through the whole sample. In other words, an ideal superconductor for which all surface effects are suppressed.

\begin{figure}
\includegraphics[width=1\columnwidth,height=0.7\linewidth]{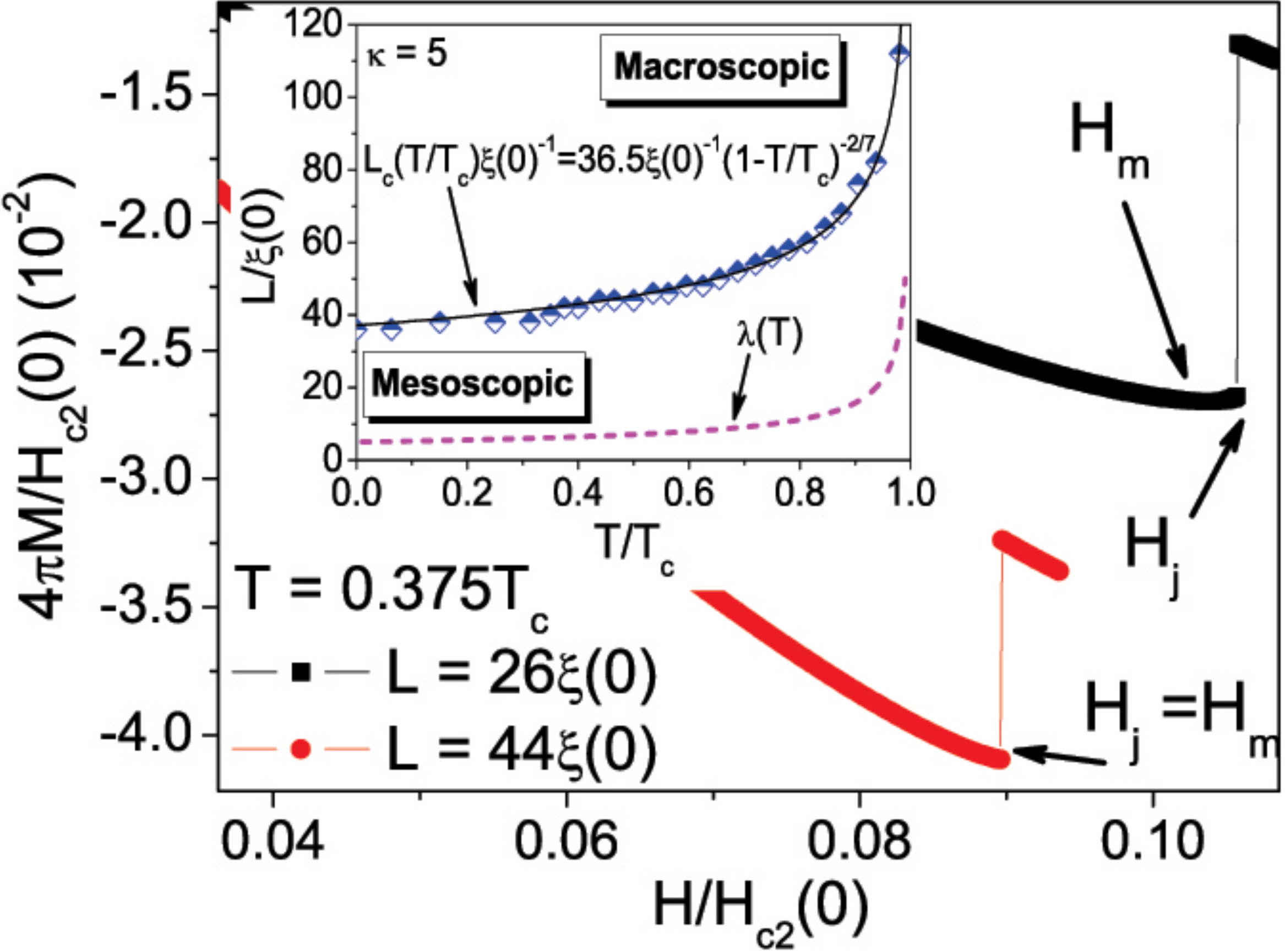}
\caption{(Color online) Magnetization versus applied magnetic field curve for superconducting squares of sizes $L=26\xi(0)$ and $44\xi(0)$ at $T=0.375 T_c$. As $L/\xi(0)$ increases, $H_j$ and $H_m$ become close one each other and when they are coincident the corresponding size is chosen as the threshold point between meso-to-macro behaviors for such temperature. The inset shows the $L_c(T)/\xi(0)$ diagram in comparison with $\lambda(T)$.}
\label{Fig1}
\end{figure}

Recently, by solving the TDGL equations for many dimensions of a square and many temperatures, the authors of reference \cite{zad2} developed a work where they built a diagram of the size of square superconducting samples versus the temperature, $L_c(T)$, as shown in the inset of Fig.~\ref{Fig1}. This diagram delimits two distinct behaviors of type II superconductors, i.e., they have shown the existence of a threshold line between mesoscopic and macroscopic superconducting behaviors. This curve is quite different of the penetration depth of the material, $\lambda (T)$, which is commonly used as the definition for the typical size of mesoscopic samples. Thus, the curve $L_c(T)$ represents the \textit{meso-to-macro crossover}, that is, a length which nearly separates the mesoscopic and macroscopic regimes described above. We must stress that this is \textit{not a phase transition}. Instead, it is a smooth change from one regime to another.

The criterion used in \citep{zad1,zad2} to obtain the meso-to-macro crossover was based on the surface barrier for the first penetration of vortices inside the superconductor. In Fig.~\ref{Fig1}, we show the magnetization curve $M(H)$ as a function of the applied field $H$ for a superconducting square of sizes $L=26\xi(0)$ and $L=44\xi(0)$, and temperature $T=0.375 T_c$. In this figure we indicate the minimum of the magnetization as $H_m$, and by $H_j$ the field which corresponds to the first nucleation of vortices. We can easily observe that, as $L/\xi(0)$ increases, $H_m$ and $H_j$ become very close to one each other. When we achieve a determined length $L/\xi(0)$ for which these two values of the applied field are equal, within a certain precision, we approach the meso-to-macro crossover. In \cite{zad2} we explain in more details the physical basis of the criterion. We then repeat the procedure for several temperatures and obtain an $L_c(T)/\xi(0)$ diagram.

In the present work, we will focus on the \textit{macro-to-bulk crossover}. The criterion which we will rely on is the initial slope of the magnetization curve in the Meissner state. In the next section we will describe this criterion in more details. We will see that as $L/\xi(0)$ increases, in the Meissner state, the curve $M(H)$ moves toward the expected bulk line $4\pi M(H)=-H$. Rigorously speaking, we would only achieve the bulk regime for $L=\infty$. Since we are only interested in a crossover, we will relax this criterion and use a slope which is slightly larger than $-1$, say, $-0.999$. Of course there is an arbitrariness in this criterion. The best choice of the inclination only could be tested experimentaly, in which the surface effects should be minimally reduced.

\section{Results and Discussion}
\label{Results}

\begin{figure}
\includegraphics[width=1\columnwidth,height=0.7\linewidth]{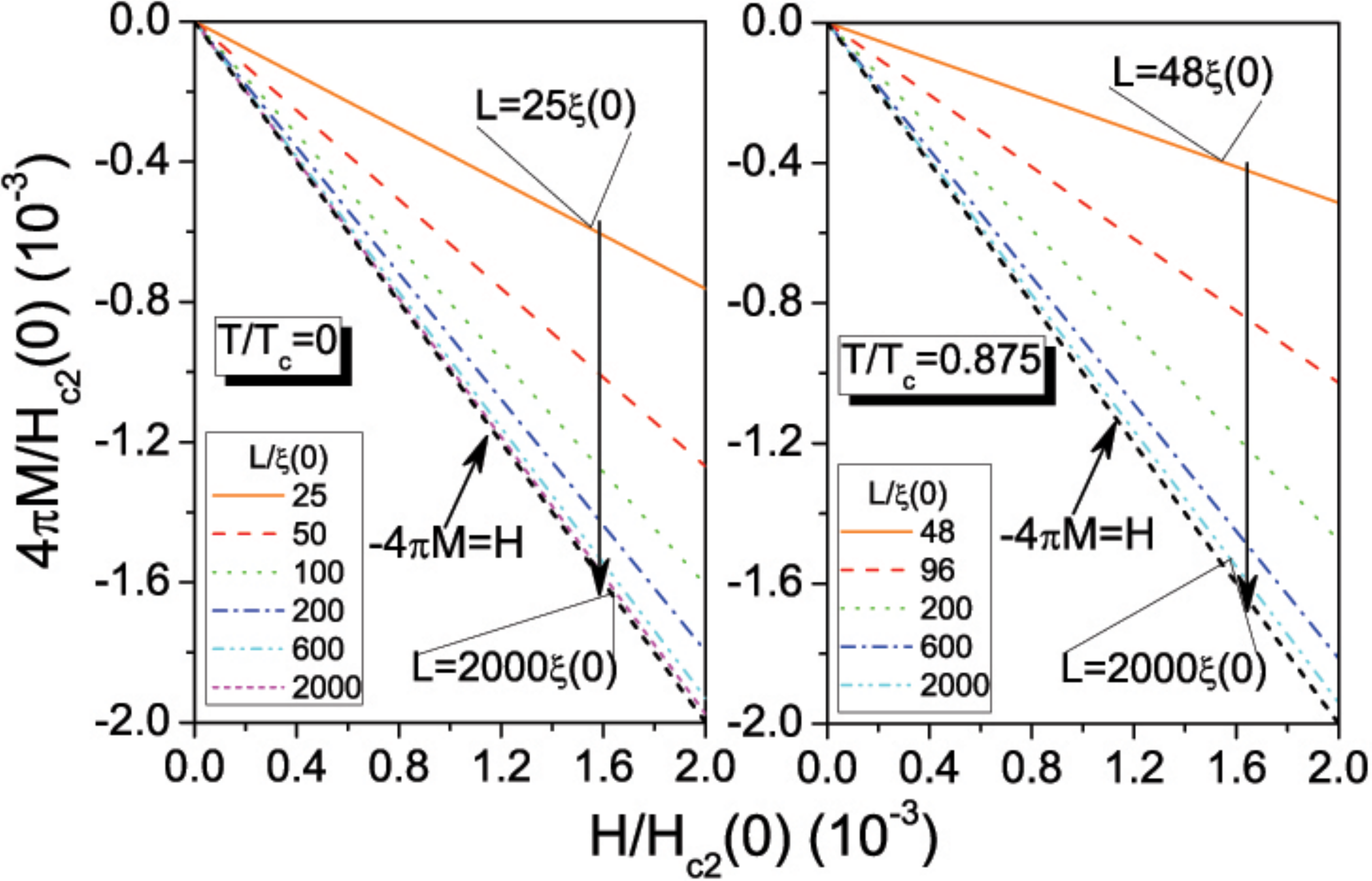}
\caption{(Color online) Magnetization as a function of the applied magnetic field for several values of the samples size for $T/T_c = 0.0$ and $0.8750$. We can note that as $L/\xi(0)$ increases, the curves approximate the Meissner line which is predicted for bulk samples.}
\label{Fig2}
\end{figure}

We simulated several samples with different lateral sizes for four values of temperature, $T/T_c = 0.0; 0.3125; 0.6500$ and $0.8750$, where $T_c$ is the critical temperature. As an illustration of the criterion used, in Fig.~\ref{Fig2} we show the $M(H)$ curves, normalized by the upper critical field at zero temperature, $H_{c2}(0)$, at $T/T_c=0.0$ and $0.8750$, for several values of $L/\xi (0)$. We can notice that, as the lateral size of the sample increases, the Meissner line moves toward the bulk curve, i.e., $4\pi M=-H$. However, even for $L=2000\xi(0)$ the curves are not entirely coincident with the bulk one, although are very closed to.

\begin{figure}
\includegraphics[width=1\columnwidth,height=0.8\linewidth]{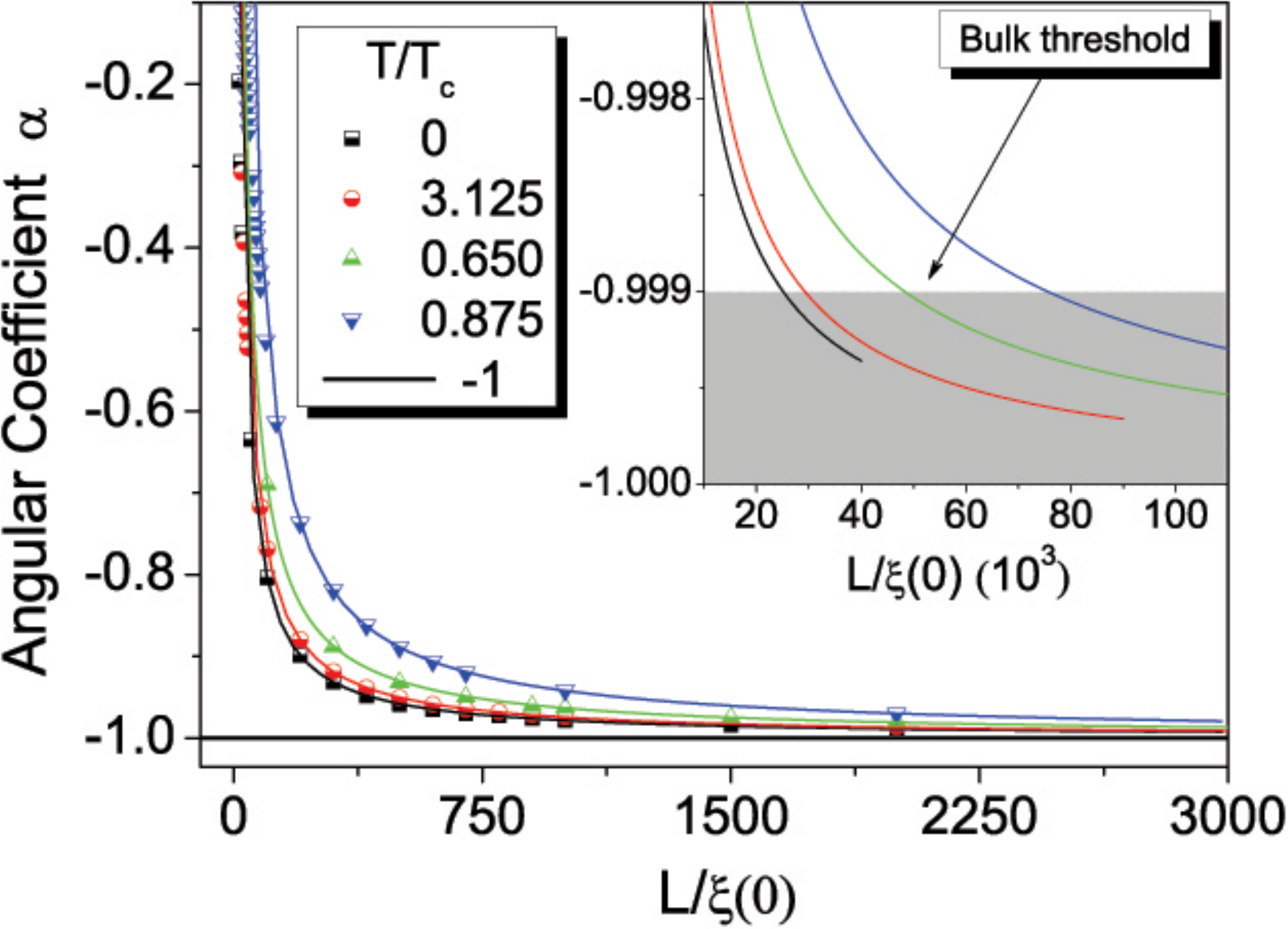}
\caption{(Color online) Angular coefficient of the $M(H)$ curves as a function of the size of the sample for $T/T_c = 0.0; 0.3125; 0.6500$ and $0.8750$. The horizontal line is the angular coefficient of bulk samples. The lines which connect the data were obtained by an exponential expression used to fit each curve.}
\label{Fig3}
\end{figure}

Running the simulations for the entire range of applied field $H$, until the destruction of superconductivity, becomes unfeasible as $L$ increases. However, since we are interested only in the slope, denoted by $\alpha (L/\xi (0))$, of the $M(H)$ curves for low values of $H$, only a few values of applied field were sufficient. Therefore, we could run our simulations for sizes until $L=2000\xi(0)$. In this way we plotted $\alpha (L/\xi (0))$ for the four studied temperatures as shown in the main panel of Fig.~\ref{Fig3}. Such curves were fitted by the exponential expression, $\alpha (L/\xi(0)) = \alpha_0 [(L/\xi(0))^{-n} e^{-L/L_0}] - 1$, where $\alpha_0$, $n$ and $L_0$ are adjustable parameters which were maintained free in such process, and the value $-1$ is the angular coefficient for bulk samples normalized by $4\pi$.

\begin{figure}
\includegraphics[width=1\columnwidth,height=0.8\linewidth]{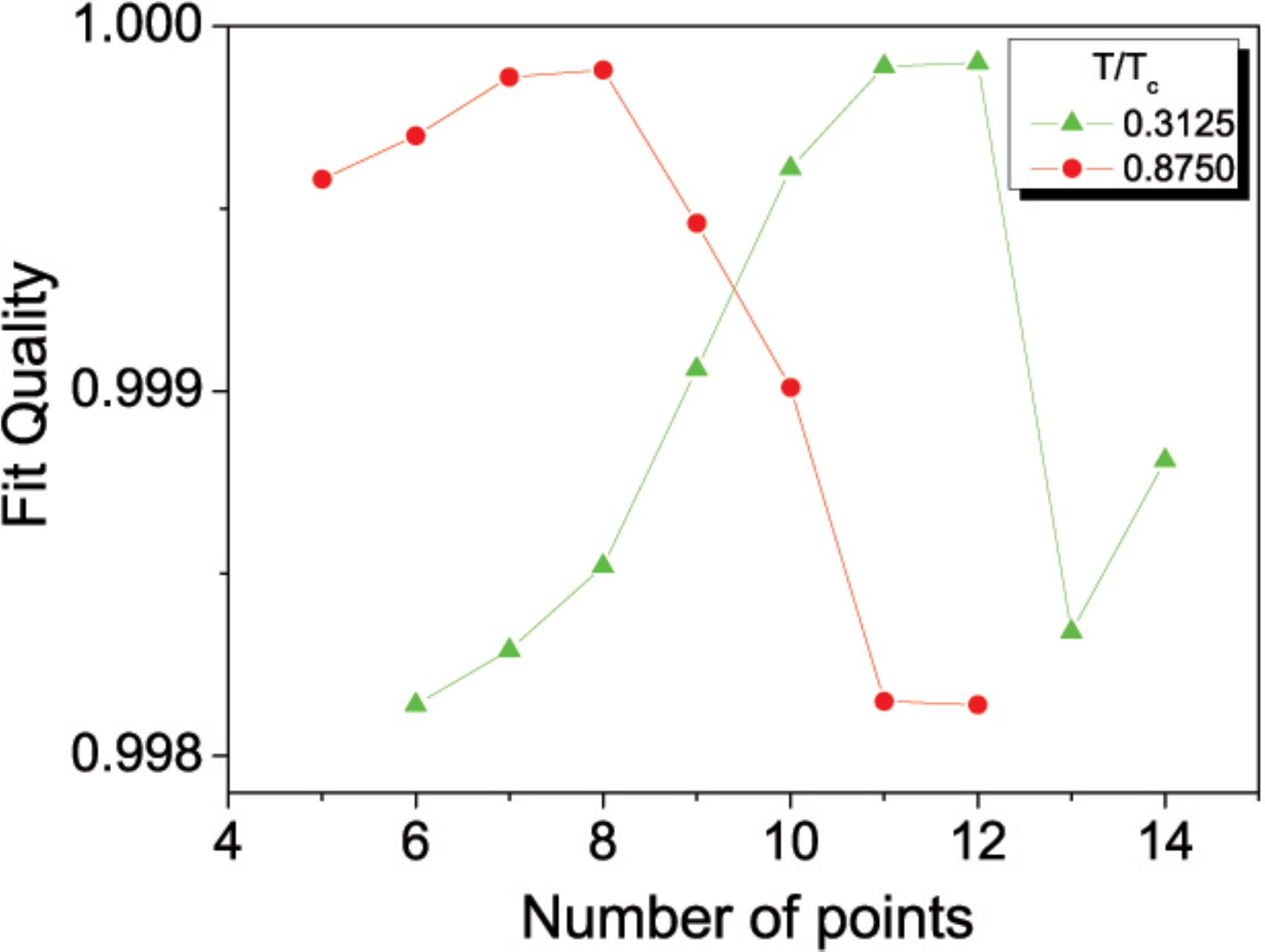}
\caption{(Color online) Fit quality as a function of the number of points considered in the fitting process for $T/T_c = 0.3125$ and $0.8750$. The number of points were counted from the last point of each curve of Fig.~\ref{Fig2} which was maintained fixed. The adjustable parameters were chosen from the best quality point.}
\label{Fig4}
\end{figure}

To choose the best parameters to fit the data of Fig.~\ref{Fig3}, we analyzed the fit quality for several amount of points. To count the number of points, we fixed the last one, i.e., $L/\xi(0)=2000$, and started from it. Fig.~\ref{Fig4} shows the curves of the fit quality for $T/T_c=0.3125$ and $0.8750$ as a function of the number of points used in each fitting process. The adjustable parameters were obtained from the best quality fit point of Fig.~\ref{Fig3} and were used to extrapolate the data until we reach the bulk region. Those fitted curves are connecting the data of Fig.~\ref{Fig3} and their extrapolation part are shown in the inset of the same figure.

\begin{figure}
\includegraphics[width=1\columnwidth,height=0.8\linewidth]{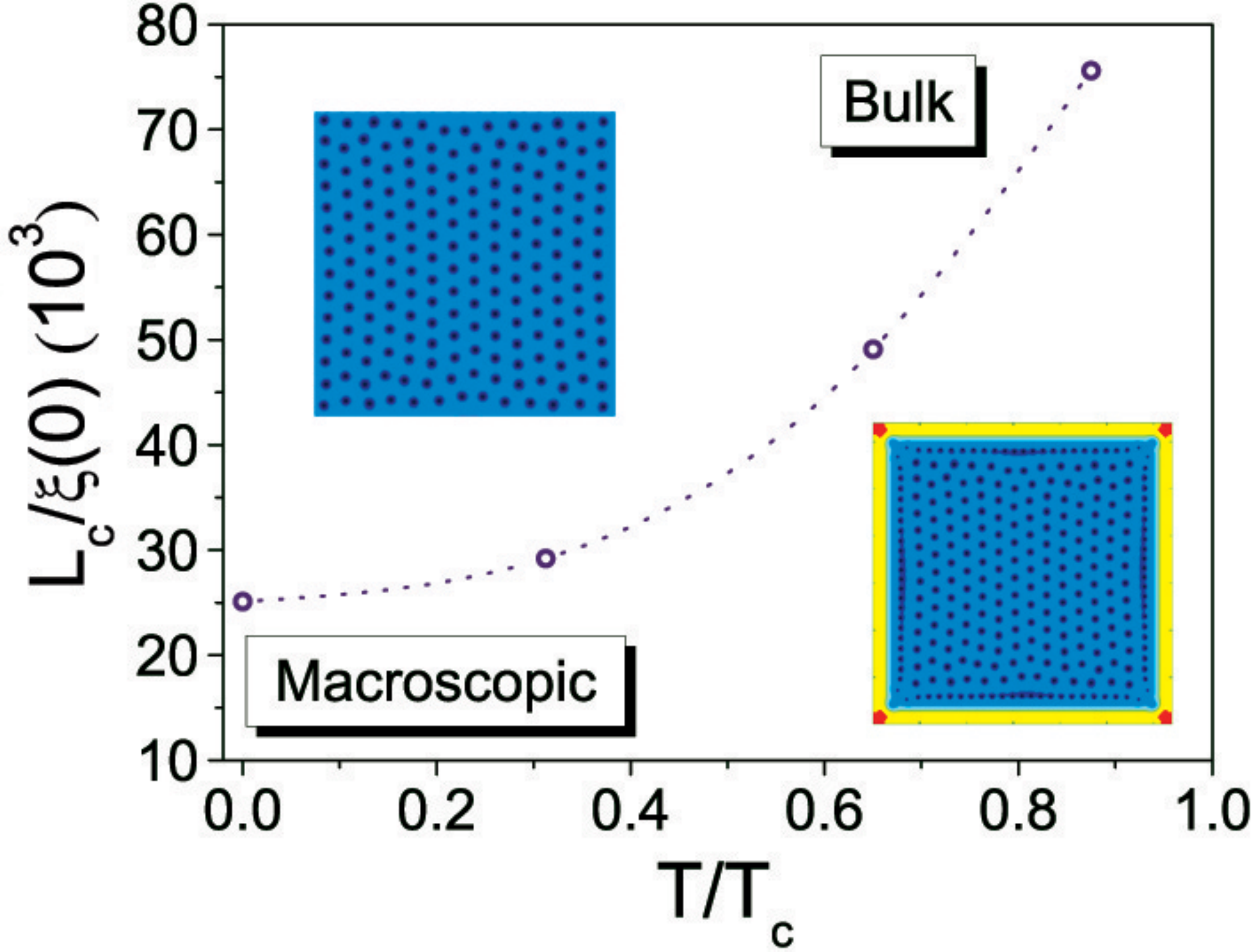}
\caption{(Color online) Sample size versus temperature diagram. The data were obtained from the point of the fitted curve which rich $0.1\% $ of the value of $\alpha (L/\xi (0))$ predicted for a bulk sample. The upper and lower inset are only representative of the macroscopic and bulk behaviors respectively.}
\label{Fig5}
\end{figure}

We considered as the criterion for the beginning of the bulk behavior, the point of the fitted curve which reached $0.1\% $ of the value of $\alpha$ predicted for bulk samples, i.e., in our case, $-1$, as shown in the inset of Fig.~\ref{Fig3}. By this analyses a diagram of the size of the sample as a function of the temperature, $L_c(T)/\xi(0)$,  was built. Such diagram gives us a reference of the threshold between the macroscopic and the bulk behaviors, as shown in Fig.~\ref{Fig5}. It is worth noticing that by a macroscopic behavior we mean the fact that the Abrikosov vortex lattice is formed in the sample, in contrast with that occurs in the mesoscopic regime \cite{zad1,zad2}, although the presence of surface superconductivity is still present, as illustrated in the lower inset of Fig.~\ref{Fig5}. On the other hand, for the macroscopic behavior, we can neglect the surface contribution, as illustrated by the upper inset of Fig.~\ref{Fig5}. In other words, the vortex configuration in the bulk regime is not influenced by surface for values of $L$ larger than critical length $L_c(T)$.

In references \cite{zad1,zad2} we have demonstrated that the meso-to-macro threshold line can be fitted as $L_c(T)=L_c(0)(1-T/T_c)^\nu$, having $L_c(0)$ and $\nu$ as fitting parameters. Since we have a few values of $L_c(T)$ we did not attempt to adjust the curve of Fig.~\ref{Fig5}, although it is very similar to Fig.~7 of \citep{zad2} where we had many more critical points. The difference in both curves is their amplitude, that is, the value of the critical length at zero temperature is very significant. In fact, from Fig.~\ref{Fig4} we obtain $L_c(0)=25105\xi(0)$ for the macro-to-bulk crossover, whereas for the meso-to-macro crossover we have $L_c(0)=36.5\xi(0)$ (see inset of Fig.~\ref{Fig1}).

\section{Conclusion}
\label{Conc}

We studied the behavior of the characteristic curves of the Meissner state for homogeneous superconducting samples with square geometry. From this investigation we observed that as the lateral size of the samples is increased, the $M(H)$ curves move toward the characteristic Meissner line of bulk materials, i.e., $ 4 \pi M = -H$. Thus, by the study of the angular coefficient of such curves we built a diagram which gives us a reference of the threshold between macroscopic and bulk behaviors. We quote the threshold sizes of $L/\xi (0)=25105$ and $L/\xi (0)=75595$ for $T/T_c=0.0$ and $T/T_c=0.875$ respectively, as examples.

\textbf{Acknowledgements}

We thank the Brazilian Agencies FAPESP, CAPES and CNPq for financial support.



\bibliographystyle{model3a-num-names}



\end{document}